\documentstyle[a4]{article}
\makeatletter
\def\siml{\mathrel{\mathpalette\gl@align<}}
\def\simg{\mathrel{\mathpalette\gl@align>}}
\def\gl@align#1#2{\lower.8ex\vbox{
\baselineskip\z@skip
 \ialign{$\m@th#1\hfill##\hfil$\crcr#2\crcr{\sim}\crcr}}}
\makeatother
\begin{document}
\def\thefootnote{\fnsymbol{footnote}}
\thispagestyle{empty}
\vspace*{2cm}
%\begin{flushright}
%\parbox{3.5cm}{
%KANAZAWA-96-01\\
%LMU-TPW 96-2
%
%January,1996}
%\end{flushright}
\vspace{13mm}
\begin{center}
{\Large\bf Quasi Yukawa Fixed Point due to 
Decoupling of 

SUSY Particles}
\vspace{1.0cm}

{\large \bf Tatsuo Kobayashi}
\footnote[1]{Alexander von Humboldt Fellow\\
\phantom{xxx}e-mail:kobayash@lswes8.ls-wess.physik.uni-munchen.de}

\vspace{.5cm}
{\it Sektion Physik, Universit\"at M\"unchen,\\
 Theresienstr. 37, D-80333 M\"unchen,Germany}

\vspace{.5cm}
{\large \bf and}\\
\vspace{.5cm}
{\large \bf Yoshio Yamagishi}
\footnote[2]{e-mail:yamagisi@hep.s.kanazawa-u.ac.jp}

\vspace{.5cm}
{\it Department of Physics, Kanazawa University\\
 Kanazawa 920-11, JAPAN}
\vspace{1.5cm}
\end{center}
\begin{abstract}
We study effects of SUSY particle decouplings on a quasi fixed point
(QFP) of Yukawa coupling.
From renormalization group analysis it is shown that 
if the SUSY breaking scale $M_S$ is large ($\simg 1$TeV), 
effects of decoupling of Higgsinos and squarks 
raise the top Yukawa QFP. 
This tendency is enhanced in most cases of non-universal SUSY breaking.
For the case of $M_S\siml 1$TeV, the decoupling of
gluinos lowers $m_t^{\rm QFP}$.
We checked some parameter dependencies for the top Yukawa QFP.
The bottom-top Yukawa unified case is also studied.
When top quark mass is measured more precisely, 
some patterns of soft mass spectra could be 
excluded if rather large initial top Yukawa coupling
is realized by underlying theory.
\end{abstract}
\newpage
\setcounter{footnote}{0}
\def\thefootnote{\arabic{footnote}}
%%%%%%%%%%%%%%%%%%%%%%%%%%%%
Supersymmetry (SUSY) \cite{Nilles} is now considered as a promising 
candidate for physics beyond the Standard Model (SM).
The minimal supersymmetric extension of the Standard Model (MSSM)
shows many successful results, i.e. the gauge coupling unification 
\cite{Amaldi}, the radiative symmetry breaking\cite{Inou}, etc.
Recently a quasi fixed point (QFP) of the top Yukawa coupling \cite{Ib-Lo}
was reconsidered by Lanzagorta and Ross\cite{Lanz-Ross}
and they pointed out that the QFP is also interesting in supersymmetric
models.
Such studies show that the QFP under the MSSM can predict the top quark 
mass closer to the experimental value\cite{CDF} than the SM.
Furthermore it is meaningful to study features of the QFP in
the framework of supersymmetric models from various viewpoints, 
e.g. effects of threshold corrections due to non-universal 
SUSY particle masses.

In general supergravity theories (SUGRAs) as well as superstring models 
lead to non-universal soft terms, i.e. non-universal soft scalar masses 
and gaugino masses \cite{Soft,non-univ}.
In Refs.\cite{unicoup,Alt-Koba}, it is shown that the effects of 
non-universal SUSY breaking on the gauge coupling unification are 
rather sizable.
In addition much work has been devoted to phenomenological 
implications of non-universality of 
SUSY particles \cite{Non-uni,softcp}.
In this paper we mainly study effects of such non-universal decoupling of 
SUSY particles on the QFP.

Usually the MSSM has been considered in the framework of $N$=1 minimal 
SUGRA which leads universal soft SUSY breaking terms.
This treatment is simple and has a powerful predictability so that
much work has been done under this framework.
However, in general, SUGRAs as well as superstring models often yield 
non-universal soft terms\cite{Soft}.
It seems that the minimal SUGRA is a special case
from the viewpoint of superstring theory.

The general forms of K\"ahler potential $K$ 
and the superpotential $W$ are written as follows,
\begin{eqnarray}
K&=&\kappa^{-2}\hat K(\Phi,\bar \Phi)+
K(\Phi,\bar \Phi)_{I{\bar J}}Q^I\bar Q^{\bar J}
+\left({1 \over 2}H(\Phi,\bar \Phi)_{IJ}Q^IQ^J+{\rm h.c.}\right)+\cdots,\\
W&=&\hat W(\Phi)+{1 \over 2}\tilde \mu(\Phi)_{IJ}Q^IQ^J+\cdots,
\end{eqnarray}
where $\kappa^2=8\pi/M_{\rm Pl}^2$ and $Q^I$ are chiral superfields.
The fields $\Phi^I$ belong to the hidden sector contributing 
the SUSY breaking. 
The ellipses stand for terms of higher orders in $Q^I$.
Using these, one can write down the scalar potential $V$ as follows,
\begin{equation}
V=\kappa^{-2}e^G[G_\alpha(G^{-1})^{\alpha \bar \beta}G_{\bar \beta}
-3\kappa^{-2}]+(\mbox{D-term}),
\label{scalar pot}
\end{equation}
where $G=K+\kappa^{-2}\log \kappa^6 |W|^2$ and the indices $\alpha$ and
$\beta$ denote $Q^I$ as well as $\Phi^m$.
If we take the flat limit $M_{\rm Pl}\rightarrow \infty$ preserving
the gravitino mass $m_{3/2}=\kappa^2e^{\hat K/2}|\hat W|$ 
fixed in Eq.(\ref{scalar pot}),
then the soft scalar masses $m_{IJ}$ 
for unnormalized fields $Q_I$ are derived as
\begin{equation}
m_{I{\bar J}}^2=m_{3/2}^2K_{I{\bar J}}
-F^m{\bar F}^{\bar n}[\partial_m\partial_{\bar n}K_{I{\bar J}}
-(\partial_{\bar n}K_{K{\bar J}})K^{K{\bar L}}(\partial_mK_{I{\bar L}})]
+\kappa^2V_0K_{I{\bar J}},
\end{equation}
where $F^m$ are F-terms of $\Phi^m$, $\partial_m$ denote
$\partial/\partial \Phi^m$ and $V_0$ is the cosmological constant.
%and hereafter gravitational coupling $\kappa$ is set to one.
The model which has non-minimal kinetic term, i.e.
$K_{I\bar J}\neq \delta_{I\bar J}$ could yield non-universal soft 
scalar masses.
In addition D-term contributions could lead to non-universal soft 
scalar masses.
In Ref.\cite{unicoup}, it is pointed out that if the non-universality
is large enough, the gauge-coupling unification scale becomes close
to the string scale $M_{\rm st}\sim 0.5\times10^{18}$ GeV 
\cite{Mst} and there is no need
for large string threshold corrections in such a case.
Various types of studies have been done about the non-universal soft SUSY
breaking\cite{Non-uni,softcp}.
It is also shown that such a large non-universality can be
realized if we consider the orbifold models with multi-moduli 
fields\cite{non-univ}.

The canonically normalized gaugino masses $M_a$ are derived through 
the following equation,
\begin{equation}
M_a={1 \over 2}({\rm Re} f_a)^{-1} F^m \partial_m f_a,
\end{equation}
where $f_a$ is a gauge kinetic function of a gauge group.
This shows that in general case the gaugino masses are also
non-universal as well as scalar masses.
The effects of gaugino mass non-universality enhances
the results of Ref.\cite{unicoup} furthermore\cite{Alt-Koba}.

Next we briefly review the QFP
\footnote{Note that this is not the Pendolton-Ross type of 
fixed points\cite{Pen-Ross}.
Their fixed point is exact ( not quasi ! ) if the SU(2) and U(1) gauge 
couplings and the bottom Yukawa coupling vanish.
However, it has been pointed out that the PR fixed point could not be 
reached since the interval of the energy scales 
between $M_X$ and  $M_W$ is too 
short to make the Yukawa coupling converge to the PR fixed point 
value\cite{Hill}.}
following the arguments of Ib\' a\~ nez and Lopez\cite{Ib-Lo}.
For the case with $Y_t \gg Y_b,Y_{\tau}$, renormalization group 
equations (RGE's) of gauge and Yukawa couplings are written as,
\begin{equation}
\frac{d g_i^2}{dt} = -\frac{b_i g_i^4}{(4\pi)}\ \ ,\ \ 
\frac{d Y_t}{dt} = Y_t\left(\sum_i r_i \tilde\alpha_i -s Y_t \right),
\label{rge}
\end{equation}
where
\[Y_t = \frac{h_t^2}{(4\pi)^2}\ \ , 
\ \ \tilde\alpha_i=\frac{g_i^2}{(4\pi)^2}.\]
Here $h_t$ and $g_i\ (i=1,2,3)$ denote 
the top Yukawa and gauge couplings,
respectively.
The coefficients $b_i,r_i$ and $s$ are some numerical constants 
which depend on a model, e.g. 
$(b_1,b_2,b_3)=(11,1,-3)$,
$(r_1,r_2,r_3)=(13/9,3,16/3)$ and $s=6$ for the MSSM;
$(b_1,b_2,b_3)=(41/6,-19/6,-7)$,
$(r_1,r_2,r_3)=(17/12,9/4,24/3)$ and $s=9/2$ for the SM.
One can solve these equations analytically and obtain 
the following results;
\begin{equation}
Y_t(t)=\frac{Y_t(0)E_1(t)}{1+sY_t(0)F_1(t)},
\label{an.so}
\end{equation}
where
\begin{eqnarray}
E_1(t)&=&
\prod_i(1 +b_i \tilde\alpha_i(0) t)^{\frac{r_i}{b_i}},\\
F_1(t)&\equiv&\int_0^tE_1(t')dt'\ \ ,\ \ t=2\log(M_X/Q).
\end{eqnarray}
Here $M_X$ is the initial scale where we set the initial value of 
the top Yukawa coupling $Y_t(0)$. 
This scale can be arbitrary but it seems natural for our purpose
to regard $M_X$ as the string scale $M_{\rm st}$.
If one takes the limit $Y_t(0) \rightarrow \infty$, Eq.(\ref{an.so})
becomes as follows,
\begin{equation}
Y_t^{\rm QFP}(t) \simeq \frac{E_1(t)}{s F_1(t)}.
\label{QFP}
\end{equation}
Note that there is no dependency on an initial value of
$Y_t(0)$ in the above formula.
This implies that $Y_t^{\rm QFP}$ can be treated 
as something like a fixed point value as long as $Y_t(0)$ is large enough.
This is the reason why we call $Y_t^{\rm QFP}$
as quasi Yukawa fixed point.
It seems necessary to study how large initial value of the top
Yukawa coupling is required in order to make the approximation
Eq.(\ref{QFP}) be realistic.
Since we obtain $F_1(t) \approx 200 \sim 300$ in Eq.(\ref{an.so}),
a deviation from the QFP Eq.(\ref{QFP}) is less than 1\% even in the case
with $Y_t(0)\sim 0.1$.
To show an applicable region of the QFP Eq.(\ref{QFP}) more explicitly,
we use Eq.(\ref{an.so}) with explicit values of $Y_t(0)$.
Even in the case with $Y_t(0)=0.1(0.01)$,
deviations from the QFP's are less than 0.3\% (2.5\%).
Therefore the QFP can give a good explanation for the value of the top
quark mass when such a initial Yukawa coupling is realized by
underlying theories like superstring theory.

One can easily obtain 
$m_t^{\rm QFP}/\sin\beta\equiv (\mbox{174GeV})\cdot 4\pi\sqrt{Y_t^{\rm QFP}} 
\sim$
205GeV in the MSSM 
($m_t^{\rm QFP}=$220GeV in the SM) from $Y_t^{\rm QFP}$
substituting the following experimentally measured values\cite{pdg},
$M_Z=91.187$GeV, $\alpha_3(M_Z)=0.118, \alpha(M_Z)=1/127.9$ and
$\sin^2\theta_W(M_Z)=0.2319$ 
into the above formulae.
Recently $m_t$ has been measured at TEVATRON\cite{CDF}:
\begin{eqnarray*}
m_t &=& 176\pm8\pm10 {\rm GeV} \ \ {\rm (CDF)}, \\
m_t &=& 199^{+19}_{-21}\pm22 {\rm GeV} \ \ \ {\rm (D0)}. \\
\end{eqnarray*}
It is obvious that $m_{t{\rm (MSSM)}}^{\rm QFP}$ is consistent with the D0
result while $m_{t{\rm (SM)}}^{\rm QFP}$ exceeds the upper limit slightly.
In addition, we can also make $m_t^{\rm QFP}$ of the MSSM consistent 
with the CDF results if we take $\sin\beta$ to be small enough.

To reduce Eq.(\ref{an.so}) for the MSSM, we have assumed that the RGE's 
of gauge and Yukawa couplings are exactly supersymmetric 
from the initial scale to $M_Z$.
This means that the SUSY breaking scale $M_S$ is just 
$M_Z$ although 
$M_S$ is, in general, treated as somewhat a higher scale than $M_Z$.
Therefore to discuss these scenario more precisely, one must 
consider effects of decoupling of SUSY particles.
If SUSY breaking is universal,
we must evaluate the QFP at $M_S$ in Eq.(\ref{QFP}), input this value of 
$Y^{\rm QFP}_t$ at $M_S$ into the initial condition of the SM RGE Eq.(\ref{rge})
and then flow $Y_t$ from $M_S$ down to $M_Z$ by the SM RGE.
Taking this prescription, we find that $m_t^{\rm QFP}$, 
the top mass predicted by the QFP,
is raised slightly if $M_S\simg 1$ TeV.
This reason is as follows.
We can write down the RGE of the top Yukawa coupling as 
\begin{equation}
\frac{dY_t}{dt}=Y_t\left(\frac{16}{3}T_{t3}\tilde\alpha_3
+3 T_{t2}\tilde\alpha_2 + \frac{13}{9}T_{t1}\tilde\alpha_1
-6 T_{tt}Y_t - T_{tb}Y_b\right).
\label{topY}
\end{equation}
Here the coefficients $T$'s are obtained as 
$T_{t3}=T_{t2}=T_{t1}=T_{tt}=T_{tb}=1$ for the MSSM and 
$T_{t3}=3/2,T_{t2}=3/4,T_{t1}= 51/52,T_{tt}=3/4$ and $T_{tb}=1/2$ 
for the SM.
In the above RGE the terms including gauge couplings 
make $Y_t$ go upward and the Yukawa term plays an opposite role
while running from the higher scale.
This is caused by the difference of the signs of these terms
\footnote{A similar situation occurs in the scenario of the radiative
symmetry breaking\cite{Inou}.
The Higgs squared mass can be negative while running from an initial scale 
to $M_Z$ since in the Higgs mass RGE the Yukawa term is dominant
against the gauge terms.}.
Note that the SM has a larger value of $T_{t3}$ and a smaller 
value of $T_{tt}$ than the MSSM.
Because of both effects of large $T_{t3}$
and small $T_{tt}$,
the SM top Yukawa RGE has a stronger tendency 
to push $Y_t$ upward during the running 
from $M_S$ to $M_Z$ than the MSSM.
Therefore if one stops the QFP Eq.(\ref{QFP}) at $M_S$ and runs $Y_t$ by the 
SM RGE from $M_S$ to $M_Z$, one obtains a larger value of $m_t^{\rm QFP}$ than
the usual QFP analysis.
We could expect that non-universal decoupling has more complicated 
effects on the evolution of the top Yukawa coupling.

To discuss such effects on the QFP, we must consider 
the decoupling in the RGE of the Yukawa coupling.
The 1-loop Yukawa RGE's including the effects of decoupling of SUSY particles 
have been presented by Lahanas and Tamvakis\cite{La-Tam}.
They parameterize the decoupling of SUSY particles by the 
step-function $\theta_\phi=\theta(Q^2 - m_\phi^2)$. 
Using this step-function approximation they
derive RGE's not only for Yukawa couplings but also gauge couplings,
A-parameters, scalar masses and gaugino masses.
These RGE's are very useful to analyze features
of the decoupling
\footnote{These RGE's correspond to the two-Higgs doublet model when 
all SUSY particles are decoupled. 
To get the ordinary SM with one Higgs doublet,
we must take into account the mixing of Higgses.
However, in the case which contains mixing of fields,
the mass-independent renormalization with the 
$\theta$- function approximation is not applicable
and a mass-dependent renormalization scheme should be taken 
instead of that.}.
We can investigate the Yukawa coupling QFP in various patterns of 
SUSY mass spectra using these RGE's.

The non-universality affects $\alpha_i(M_{\rm st})$ and 
the running of gauge and Yukawa
couplings below $M_S$.
These are crucial for the determination of the QFP value
so that it is important to study effects of 
non-universal soft SUSY breaking on the QFP. 
Our procedure is following. 
Firstly, we determine $\alpha_3(M_Z),\alpha_2(M_Z)$ and $\alpha_1(M_Z)$
by experimental value and let them run from $M_Z$ to $M_{\rm st}$.
Up to the SUSY breaking scale $M_S$ we use the corresponding 
RGE's (\ref{rge})
for each non-universal case following Refs.\cite{unicoup,Alt-Koba}.
Then we turn on all contributions from SUSY particles at $M_S$, hence
the flow of those couplings obeys the usual MSSM RGE's
from $M_S$ to $M_{\rm st}$.
After evaluating $\alpha_i(M_{\rm st})$'s we input them and $M_S$
to the QFP formula (\ref{QFP})
to obtain the value of the top Yukawa
coupling at the SUSY breaking scale $M_S$. Finally we let 
gauge and top Yukawa couplings flow down from $M_S$ to $m_t$ 
by the corresponding RGE's and evaluate $m_t^{\rm QFP}$
for each non-universal case.
Hereafter we consider eight patterns of non-universalities 
shown in Ref.\cite{Alt-Koba}.
We review them in Table 1 with $\beta$-coefficients $b_i\ ,\ i=1,2,3$ 
for each case.
Although Ref.\cite{Alt-Koba} gives ten cases, we take eight out of ten 
since Case II and V, Case C and D indicate almost same
behavior respectively in the following analysis.
For each case the coefficients of the RGE's in Eq.(\ref{topY}) $T$'s are 
given in Table 2 and 3.
We follow the notation of Ref.\cite{La-Tam}.
For the evaluation of $m_t^{\rm QFP}$,
we use the following relation between the running top quark 
mass and the pole mass\cite{pole};
\begin{equation}
m_t=\bar m_t(m_t) \left[1+\frac{4\bar\alpha_3(m_t)}{3\pi}
+O(\alpha_3^2)\right],
\label{pole}
\end{equation}
where $\bar m_t(m_t)$ and $\bar \alpha_3(m_t)$ denote the running
top quark mass and the running SU(3) gauge coupling at $m_t$ 
respectively.

Firstly we concentrate on the $Y_t \gg Y_b,Y_{\tau}$ case.
We assume $\tan\beta=2$ in all patterns of
non-universalities, for simplicity. If $Y_t \gg Y_b,Y_{\tau}$, 
a value of $\tan\beta$ can not be so large.
Because a large value of $\tan\beta$ leads to a large value of $Y_b$ 
in order to realize the bottom quark mass.
However, the $\tan\beta$ dependence on $m_t^{\rm QFP}$ is very large.
For example, if $\tan\beta$ is changed from 2 to 1.5,
$m_t^{\rm QFP}$ becomes 168 GeV from 181 GeV.
Hereafter we take $\alpha_3=0.118$ and $M_X=M_{\rm st}
\equiv 5.0\times 10^{17}$GeV.

The results are shown in Figures 1-2.
Figure 1 shows the $m_t^{\rm QFP}$ dependencies on $M_S$ for each
non-universal case.
For the present we assume that all Higgsinos are decoupled at 
$M_S$.
Case 0 is the usual $M_S=M_Z$ prescription taken in 
the analysis of Ref.\cite{Ib-Lo,Lanz-Ross}.
Case I corresponds to the universal SUSY breaking case
which decouples all SUSY particles at $M_S$.
All cases raise $m_t^{\rm QFP}$ 
in the region of $M_S\simg 1$TeV more than case 0.
The cases with the same value of $b_3$ behave similarly.
This is because of the fact that 
the coefficient $T_{tt}$ is completely identical in all cases.
In such cases the difference of $m_t^{\rm QFP}$
mainly depends on $\alpha_3$.
The running of $\alpha_3$ is determined 
by $b_3$, so that
the cases with the same $b_3$ value tend to behave similarly.

One can expect from the Yukawa RGE's that
the decoupling of Higgsinos is quite effective for this kind of analysis.
We also consider the cases with the light Higgsinos and the results are
shown in Figure 2.
From this figure one can see that in a higher $M_S$ region
$m_t^{\rm QFP}$ is rather separated from each other compared with 
the previous cases with the heavy Higgsinos.
As we shown before, this is due to the sameness of $T_{tt}$.
However, the light Higgsinos yield the difference in $T_{tt}$ 
and cause such a separation of $m_t^{\rm QFP}$.
\footnote{This analysis seems to be unacceptable
for the explaination of the separation between Case II and B.
In these cases the differences appear in SU(2) coefficients
($T_{t2},b'_2$). At higher energy scale, the gauge coupling of
SU(2) is effective as much as SU(3).
Therefore it seems plausible
to consider that the separation of the lines 
in Figure 2 is also triggered by
the descrepancy of the SU(2) coefficients.}
This effect is significant in Case IV.
In this case all squarks are light.
This effect appears in the fact that Case IV has the lowest $T_{t3}$ 
value as $T_{t3}=1$.
As shown before, the SU(3) gauge interaction raises the $m_t^{\rm QFP}$
as the renormalization scale is going down.
However, for Case IV, such RG effects are not operative
because of small $T_{t3}$.
From the above results we conclude that
light squarks and Higgsinos are favorable 
in order to lower $m_t^{\rm QFP}$ for $M_S \simg 1$TeV region.

When $M_S \siml $ 1 TeV, the situation becomes different.
In this region, Case I and C generate a smaller value of $m_t^{\rm QFP}$ 
than the usual QFP analysis, Case 0.
When $M_S$ is small, the RG effects in Yukawa coupling
do not matter because of such a short running
interval between $M_S$ and $M_Z$.
The common property of Case I and C is the decoupling of 
gluinos.
If gluinos are decoupled,
the $\beta$ coefficient for the SU(3) gauge coupling $b_3$
is decreased and $\alpha_3(m_t)$ become small due to RG effects. 
At the renormalization scale close to $m_t$,
the correction term of $O(\alpha_3(m_t))$
in Eq.(\ref{pole}) strongly influences 
$m_t^{\rm QFP}$.
One can read off from Table 1 and Figures 1-2 that $m_t^{\rm QFP}$ 
of the cases with the same $b_3$ converge 
to the same points respectively at $M_S \simeq 200$ GeV .

The large ambiguity of the experimental value of $\alpha_3(M_Z)$
is still a big problem for phenomenologists.
The QFP is also affected by the value of
$\alpha_3(M_Z)$.
For $\alpha_3(M_Z)=0.110(0.130)$ and $M_S=200$GeV, 
Cases 0, II and III with the heavy Higgsinos lead to 
$m_t^{\rm QFP}=174$(183), 176(185) and 177(186)GeV, respectively.
We also obtain $5\sim6$ \% difference for $m_t^{\rm QFP}$ against 
$\alpha_3(M_Z)=0.11\sim 0.13 $ in other cases.
A smaller value of $\alpha^{-1}_3(M_Z)$ lowers $m_t^{\rm QFP}$.

We also examine the dependence on a starting scale $M_X$.
For $M_X=10^{16}$GeV($10^{19}$GeV) and $M_S=10$TeV, Cases 0, II and III 
with the heavy Higgsinos provide $m_t^{\rm QFP}=180$(176), 182(178) and 
183(179)GeV, respectively. 
It is obvious that a higher starting scale suppresses
the value of $m_t^{\rm QFP}$.
Below $M_X$ the top Yukawa coupling tends to
decrease monotonically as the renormalization scale is going down. 
Therefore a large interval of renormalization scale due to the higher
$M_X$ lowers the QFP furthermore
\footnote{
This situation looks like the case of triviality bound of the Higgs mass
\cite{triv}.
According to the argument of triviality, 
the Higgs mass bound which is determined by the Higgs four-point coupling
decreases if cutoff $\Lambda$ increases.
}.

Next we consider the Yukawa-unified case, $Y_t\simeq Y_b$.
In this case, the formula of the top Yukawa QFP is changed into\cite{copw}
\footnote{In the derivation of Eq.(\ref{unif}),
we assume $Y_t=Y_b$ in all over the range between $M_X$ and $M_Z$.
However, this treatment is not correct because RGE coefficients
of the top Yukawa is different from those for the bottom Yukawa even 
in the MSSM, so that the two evolutions are different from each other.
Taking into account these effects, we solve RGE's
numerically and find that the results which are shown 
in Figure 3 are entirely raised by 2$\sim$3 GeV.}
\begin{equation}
Y_t(t)\simeq\frac{E_1(t)}{7 F_1(t)}.
\label{unif}
\end{equation}
Similar analysis can be done for the bottom Yukawa and we obtain
the QFP for the bottom Yukawa as follows,
\begin{equation}
Y_b(t)\simeq\frac{E_2(t)}{7 F_2(t)},
\end{equation}
where
\begin{eqnarray}
E_2(t)&\equiv&(1+b_3 \tilde\alpha_3(0) t)^{16/3 b3}
(1+b_2 \tilde\alpha_2(0) t)^{3/b2}
(1+b_1 \tilde\alpha_1(0) t)^{7/9 b1},\\
F_2(t)&\equiv&\int_0^tE_2(t')dt.
\end{eqnarray}
Below $M_S$ we use the following RGE for bottom Yukawa coupling;
\begin{equation}
\frac{dY_b}{dt}=Y_b\left(\frac{16}{3}T_{b3}\tilde\alpha_3
+3 T_{b2}\tilde\alpha_2 + \frac{7}{9}T_{b1}\tilde\alpha_1
-6 T_{bb}Y_b - T_{bt}Y_b\right).
\end{equation}
The coefficients $T$'s are shown in Tables 2 and 3.
In this case there is a constraint from the bottom quark mass.
Here we take the running bottom quark mass $m_b(m_b)$ 
as 4.1GeV
\footnote{The ambiguity of the bottom mass is rather large as 
one of $\alpha_3$. However, we check that our results are not so
sensitive to the value of $m_b$ as long as 4.1GeV$\le m_b \le$4.5GeV.}.
Consequently we obtain $\tan\beta\simeq 65$.
This value is almost common to all of the cases we used.
In the interval from $M_Z$ to $m_b$ the running of the bottom 
Yukawa coupling should obey the RGE including only QCD and electromagnetic 
gauge interaction.
Here we simply assume $m_b(M_Z)=m_b(m_b)/\eta_b$ 
where $\eta_b\simeq 1.437+0.075[\alpha_3(M_Z)-0.115]/0.01$\cite{Hall}.

We investigate the top Yukawa QFP similarly as the $h_t\gg h_b$ case and 
the results are shown in Figure 3. 
The overall tendency of results is similar to the previous cases.
It seems strange that even for a quite large $\tan\beta \ (\ \simg 60 )$,
$m_t^{\rm QFP}$ is not so large compared to the previous cases.
This fact is due to the presence of the contribution from 
the bottom Yukawa.
The top Yukawa QFP becomes smaller than in the previous cases since 
the factor $s$ in the denominator of Eq.(\ref{QFP}) 
is larger than in the $Y_t \gg Y_b$ case.
In addition, the top Yukawa RGE below $M_S$ has an additional term
from the bottom Yukawa and this term contributes to lower $h_t$ furthermore.

Finally we summarize our results.
The quasi fixed point of the top Yukawa coupling is investigated 
from various viewpoints.
If one takes into account the universal decoupling
of SUSY particles, $m_t^{\rm QFP}$ is raised
slightly when the SUSY breaking scale $M_S\simg 1$TeV.
This situation is enhanced in most of the cases with non-universal soft SUSY
breaking terms.
For the case of $M_S \siml 1$TeV, the decoupling of gluinos is
crucial for the determination of $m_t^{\rm QFP}$.
The QFP is sensitive about the experimental value of $\alpha_3$.
In most cases the QFP predicts rather large values of $m_t$.

The QFP could provide a good reason for the value of the top quark mass
if a high-energy theory like superstring theory 
or some gauge-Yukawa unified models\footnote{See e.g. Refs\cite{kubo}.}
can give explanation 
about a suitable initial Yukawa coupling.
From our analysis 
the following prescription may lower $m_t^{\rm QFP}$ enough to reconcile
with the CDF results:
1. a lower $\tan\beta$,
2. a lower SUSY breaking scale $M_S \siml 1 TeV$, 
3. heavy gluinos ($m_{\tilde{g}} \simg M_S$),
4. a lower value of $\alpha_3(M_Z)$,
5. a higher start point $(\sim M_{\rm Pl})$.
If $M_S \simg 1$ TeV,
the decoupling of squarks and Higgsinos raises
the QFP furthermore. 
In order to lower the QFP,
squarks might be light enough.
This seems unfavorable for the experimental constraints 
from electronic dipole moment of neutron (EDMN) \cite{Ki-Oshi}.
However it was pointed out that if small $\mu,\tan\beta$ and $M_2$
are realized simultaneously, 
the contribution from a dangerous soft CP-violating phase 
can be suppressed successfully\cite{softcp} and the EDMN need not
to be so large.
If the top quark mass as well as $\alpha_3$ is measured more precisely 
in future, our results become more serious.
For example the CDF and D0 results provide $m_t \sim 181$GeV as the 
mean value.
If this value included a very small error, some cases with smaller $M_S$ 
could be ruled out.

{\bf Acknowledgment}\\
The authors would like to thank R.~Altendorfer, K.~Inoue, J.~Kubo, 
N.~Polonsky and D.~Suematsu for useful discussions.

\newpage

\section*{Table 1}

The patterns of non-universal soft SUSY breaking and 
corresponding $\beta$ coefficients below $M_S$.
The capital letter in the second and third column denotes squark or
slepton and $\lambda_i$ express the gaugino.
Particles in second column are assumed to be heavy and decouple at $M_S$.
The third column is devoted to the light SUSY particles which remain 
in the scale below $M_S$.
The $b'_2$ and $b'_1$ are the SU(2) and U(1) $\beta$ coefficients 
respectively for the case of light Higgsinos.
The decoupling of Higgsinos does not affect $b_3$.

\vspace{1cm}
\hspace{-.7cm}
\begin{tabular}{|l|c|c|c|c|c|} \hline
{\normalsize  Case} 
& { $\sim M_S$} &{ $\sim M_Z$} & $b_3$ & $b_2,b_1$ 
& $b'_2,b'_1$ \\ \hline
{  I} &{ $ \tilde Q,\tilde U,\tilde D,\tilde L,\tilde
E,\lambda_3,\lambda_2,\lambda_1$} &
{\rm  (Universal)} &{ $-7$ }&{ $-3,7$ }&{ $-7/3,23/3$ } \\ \hline
{  II}&{ $\tilde Q,\tilde U,\tilde D,\tilde
L$}&{ $\tilde E,\lambda_3,\lambda_2,\lambda_1$}
&{ $-5$ }&{$-5/3,8$ }&{ $-1,26/3$ }\\ \hline 
{  III}&{ $\tilde Q,\tilde
L$}&{ $\tilde U,\tilde D,\tilde E,\lambda_3,\lambda_2,\lambda_1$}
&{ $-4$ }&{ $-5/3,29/3$ }&{ $-1,31/3$ }\\ \hline
{  IV}&{ $\tilde L$}&{
$\tilde Q,\tilde U,\tilde D,\tilde E,\lambda_3,\lambda_2,\lambda_1$}
&{ $-3$ }&{ $-1/6,59/6$ }&{ $1/2,21/2$ }\\ \hline
%{\it  V}&{ $\tilde Q,\tilde U,\tilde
%D$}&{ $\tilde
%L,\tilde E\lambda_3,\lambda_2,\lambda_1
%$}&{ 17/2 }&{ -7/6 }&{ -5 }&{ 55/6 }&{ -1/2 }&{ -5 }\\ \hline
{  VI}&{ $\tilde U,\tilde D$} &
{ $ \tilde Q,\tilde L,\tilde E,\lambda_3,\lambda_2,\lambda_1$}
&{ $-4$ }&{ $1/3,26/3$ }&{ $1,28/3$ }\\ \hline
{  A}&{ $\tilde Q,\tilde L,\lambda_2$} &
{ $ \tilde U,\tilde D,\tilde E,\lambda_3,\lambda_1$}
&{ $-4$ }&{ $-3,29/3$ }&{ $-7/3,31/3$ }\\ \hline
{  B}&{ $\tilde Q,\tilde U,\tilde D,\tilde
L,\lambda_2$}&{ $\tilde E,\lambda_3,\lambda_1$}
&{ $-5$ }&{ $-3,8$ }&{ $26/3,-7/3$ }\\ \hline 
{  C}&{ $\tilde Q,\tilde U,\tilde
D,\lambda_3$}&{ $\tilde L,\tilde E,\lambda_2,\lambda_1$}
&{ $-7$ }&{ $-7/6,17/2$ }&{ $-1/2,55/6$ }\\ \hline 
%{\it  D}&{ $\tilde Q,\tilde U,\tilde
%D,\tilde L,\lambda_3$}&{ $\tilde E,\lambda_2,\lambda_1$}&{ 8 }&{ -5/3
%}&{ -7 }&{ 26/3 }&{ -1 }&{ -7 }\\ \hline 
\end{tabular}

\newpage

\section*{Table 2}

The RGE coefficients of top and bottom Yukawa couplings
for each non-universal case.
These expressions follow the notation of Ref.\cite{La-Tam}.
In these cases all Higgsinos are assumed to be decoupled.

\vspace{1cm}
%\hspace{-1cm}
\begin{tabular}{|c||c|c|c|c|c||c|c|c|c|c|} \hline
Case& $T_{t3}$ & $T_{t2}$ &$T_{t1}$ &$T_{tt}$
& $T_{tb}$ &$T_{b3}$ & $T_{b2}$ &$T_{b1}$ &$T_{bt}$ &$T_{bb}$\\ \hline
{ I   }&{ 3/2 }&{ 3/4 }&{ 51/52 }&{ 3/4 }&{ 1/2 }&
{ 3/2 }&{ 3/4 }&{ 15/28 }&{ 1/2 }&{ 3/4 } \\ \hline
{ II  }&{ 3/2 }&{ 3/4 }&{ 51/52 }&{ 3/4 }&{ 1/2 }&
{ 3/2 }&{ 3/4 }&{ 15/28 }&{ 1/2 }&{ 3/4 } \\ \hline
{ III }&{ 5/4 }&{ 3/4 }&{ 35/52 }&{ 3/4 }&{ 1/2 }&
{ 5/4 }&{ 3/4 }&{ 11/28 }&{ 1/2 }&{ 3/4 } \\ \hline
{ IV  }&{ 1 }&{ 1/2 }&{ 17/26 }&{ 3/4 }&{ 1/2 }&
{ 1 }&{ 1/2 }&{ 5/14 }&{ 1/2 }&{ 3/4 } \\ \hline
%{  V  }&{ 17/2 }&{ -7/6 }&{ -5 }&{ 3/2 }&{ 3/4 }&{ 51/52 }&{ 3/4 }&{ 1/2 }&
%{ 3/2 }&{ 3/4 }&{ 15/28 }&{ 1/2 }&{ 3/4 } \\ \hline
{ VI  }&{ 5/4 }&{ 1/2 }&{ 25/26 }&{ 3/4 }&{ 1/2 }&
{ 5/4 }&{ 1/2 }&{ 1/2 }&{ 1/2 }&{ 3/4 } \\ \hline
{ A   }&{ 5/4 }&{ 3/4 }&{ 51/52 }&{ 3/4 }&{ 1/2 }&
{ 3/2 }&{ 3/4 }&{ 11/28 }&{ 1/2 }&{ 3/4 } \\ \hline
{ B   }&{ 3/2 }&{ 3/4 }&{ 51/52 }&{ 3/4 }&{ 1/2 }&
{ 3/2 }&{ 3/4 }&{ 15/28 }&{ 1/2 }&{ 3/4 } \\ \hline
{ C   }&{ 3/2 }&{ 3/4 }&{ 51/52 }&{ 3/4 }&{ 1/2 }&
{ 3/2 }&{ 3/4 }&{ 15/28 }&{ 1/2 }&{ 3/4 } \\ \hline
%{ D   }&{ 8 }&{ -5/3 }&{ -7 }&{ 3/2 }&{ 3/4 }&{ 51/52 }&{ 3/4 }&{ 1/2 }&
%{ 3/2 }&{ 3/4 }&{ 15/28 }&{ 1/2 }&{ 3/4 } \\ \hline
\end{tabular}

\section*{Table 3}

The RGE coefficients of top and bottom Yukawa couplings
for each non-universal case.
In these cases all Higgsinos are assumed to be light.

\vspace{1cm}
%\hspace{-1cm}
\begin{tabular}{|c||c|c|c|c|c||c|c|c|c|c|} \hline
Case & $T_{t3}$ & $T_{t2}$ &$T_{t1}$ &$T_{tt}$
& $T_{tb}$ &$T_{b3}$ & $T_{b2}$ & $T_{b1}$ & $T_{bt}$ & $T_{bb}$ \\ \hline
{ I   }&{ 3/2 }&{ 3/4 }&{ 51/52 }&{ 3/4 }&{ 1/2 }&
{ 3/2 }&{ 3/4 }&{ 15/28 }&{ 1/2 }&{ 3/4 } \\ \hline
{ II  }&{ 3/2 }&{ 1/4 }&{ 33/52 }&{ 3/4 }&{ 1/2 }&
{ 3/2 }&{ 1/4 }&{ -3/28 }&{ 1/2 }&{ 3/4 } \\ \hline
{ III }&{ 5/4 }&{ 1/4 }&{ 65/52 }&{ 5/6 }&{ 1 }&
{ 5/4 }&{ 1/4 }&{ 17/28 }&{ 1 }&{ 5/6 } \\ \hline
{ IV  }&{ 1 }&{ 1 }&{ 1 }&{ 1 }&{ 1 }&
{ 1 }&{ 1 }&{ 1 }&{ 1 }&{ 1 } \\ \hline
%{  V  }&{ 55/6 }&{ -1/2 }&{ -5 }&{ 3/2 }&{ 1/4 }&{ 33/52 }&{ 3/4 }&{ 1/2 }&
%{ 3/2 }&{ 1/4 }&{ -3/28 }&{ 1/2 }&{ 3/4 } \\ \hline
{ VI  }&{ 5/4 }&{ 1 }&{ 5/13 }&{ 11/12 }&{ 1/2 }&
{ 5/4 }&{ 1 }&{ 2/7 }&{ 1/2 }&{ 11/12 } \\ \hline
{ A   }&{ 5/4 }&{ 3/4 }&{ 65/52 }&{ 5/6 }&{ 1 }&
{ 5/4 }&{ 3/4 }&{ 17/28 }&{ 1 }&{ 5/6 } \\ \hline
{ B   }&{ 3/2 }&{ 3/4 }&{ 33/52 }&{ 3/4 }&{ 1/2 }&
{ 3/2 }&{ 3/4 }&{ -3/28 }&{ 1/2 }&{ 3/4 } \\ \hline
{ C   }&{ 3/2 }&{ 1/4 }&{ 33/52 }&{ 3/4 }&{ 1/2 }&
{ 3/2 }&{ 1/4 }&{ -3/28 }&{ 1/2 }&{ 3/4 } \\ \hline
%{ D   }&{ 26/3 }&{ -1 }&{ -7 }&{ 3/2 }&{ 1/4 }&{ 33/52 }&{ 3/4 }&{ 1/2 }&
%{ 3/2 }&{ 1/4 }&{ -3/28 }&{ 1/2 }&{ 3/4 } \\ \hline
\end{tabular}

\newpage
\section*{Figure Captions}
\newcounter{fig}
\begin{list}%
{\large\bf Fig.\arabic{fig}}{\usecounter{fig}
        \setlength{\rightmargin}{\leftmargin}
\setlength{\listparindent}{0cm}
\setlength{\labelsep}{.5cm}\noindent}

\item The value of $m_t^{\rm QFP}$ corresponding 
to each SUSY breaking scale $M_S$. 
In this case Higgsinos are assumed to be heavy.

\item The value of $m_t^{\rm QFP}$ corresponding 
to each SUSY breaking scale $M_S$. 
In this case all Higgsinos are assumed to be light.

\item The value of $m_t^{\rm QFP}$ corresponding 
to each SUSY breaking scale $M_S$ for the case of Yukawa unification. 
In this case all Higgsinos are assumed to be light.
\end{list}

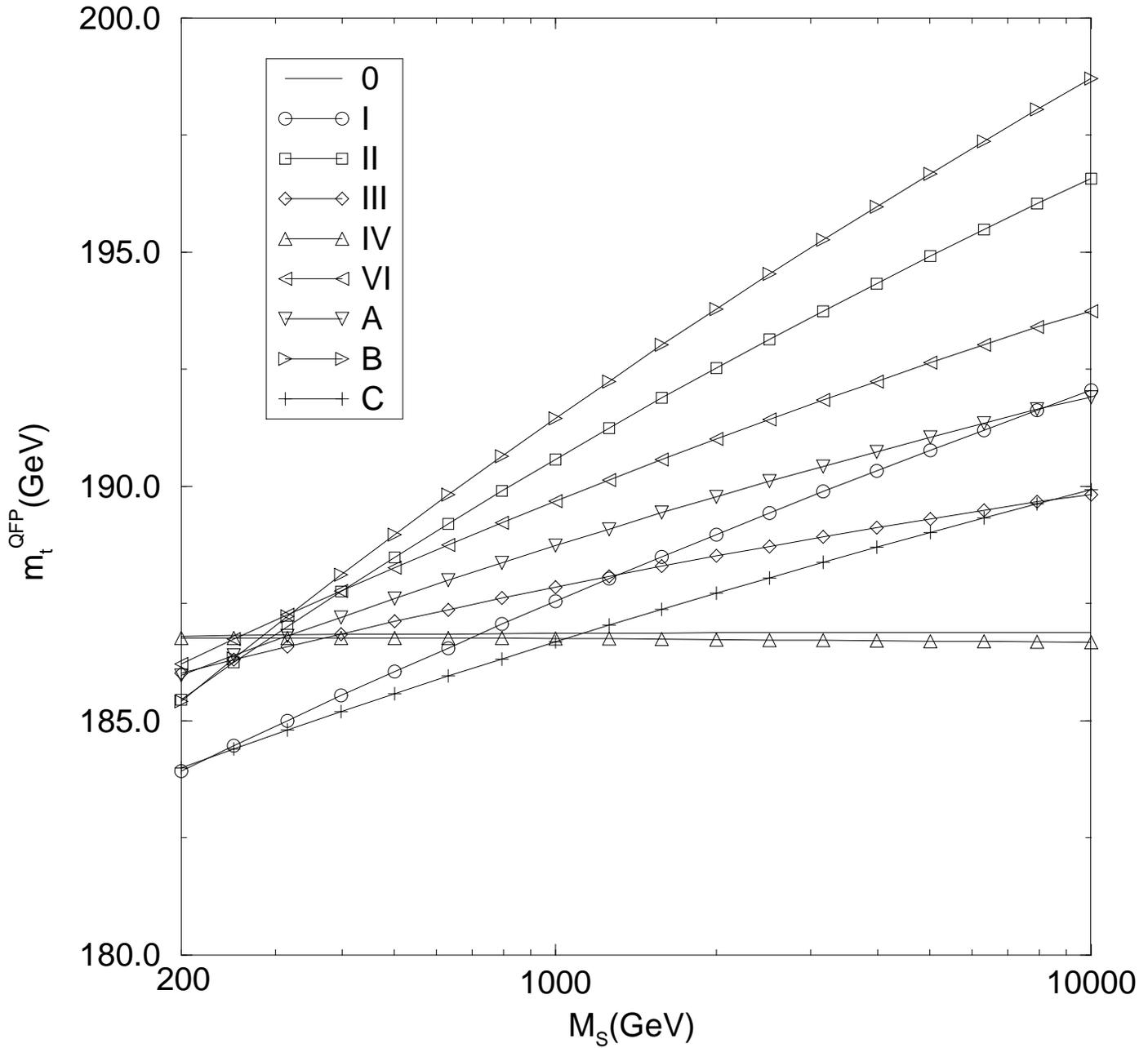
\end{document}